\documentstyle[twocolumn,aps]{revtex}
\begin{document}

\noindent{\bf Li et al. reply:}  
In their Comment\cite{goldmancomment} on our recent Letter 
\cite{LHW}, Bhattacharya et al.  note that for the geometry 
used in their experiment\cite{goldman} to try to detect the 
modes supported by the Meissner state of a $d$-wave 
superconductor predicted by $\breve{\rm Z}$uti\'c and Valls \cite{ZV}, 
a field of first vortex penetration $H_{c1}^*$ of 300 Oe 
was measured.  They then argue that the crossover field
$H^*$ separating the regime dominated by nonlocal 
electrodynamics and a local regime is of order only 20 Oe, 
suggesting that nonlocal effects can be neglected for their 
geometry for most of the intermediate field range 
$H^* < H < H^*_{c1}$. Instead, they attribute their failure 
to observe these modes to the absence of gap nodes due to 
an out-of-phase admixture of a secondary order parameter 
component. Here
we question whether the theory of $\breve{\rm Z}$uti\'c and Valls can be 
applied in this intermediate range, argue that a $d+is$ state 
in YBa$_2$Cu$_3$O$_{6.95}$ (YBCO) is unsupported by other types of 
experimental evidence, 
and continue to maintain that the nonlinear Meissner effect
is unlikely to be observed in any of its manifestations,
due in part to nonlocal effects.

In our paper, we pointed out simply that the nonlocal crossover 
field $H^*\simeq {\Phi_0 \over \pi\lambda^2_0} {\xi_{0c} \over \xi_0} $
($\lambda_0$ is the penetration depth for 
supercurrents in the plane, and $\xi_0$ and $\xi_{0c}$ are
the coherent lengths in the plane and along the $\hat{c}$ axis,
respectively), for the geometry in question, with a dominant 
(001) surface and $H\parallel {\hat a},{\hat b}$, is 
generically of the same order of magnitude as the field below 
which the Meissner state is thermodynamically stable, 
$H_{c1}\simeq {\Phi_0\over 4\pi\lambda_0\lambda_{0c}} {\rm ln}
\bar{\kappa} $, where $\lambda_{0c}$ is the penetration depth for 
supercurrents along the $\hat c$ axis, 
and $\bar{\kappa}=\sqrt{\kappa_0 \kappa_{0c}} $
with $\kappa_0 $ and $\kappa_{0c}$ the plane and the $\hat{c}$-axis 
Ginzburg-Landau parameters, respectively. Bhattacharya et al. 
argue \cite{goldmancomment,goldman} that the experimentally 
measured field of first vortex penetration $H^*_{c1}$ is as 
much as an order of magnitude higher.
With this problem in mind, we in fact alluded
to precisely this difficulty of determining the correct lower
critical field in our paper \cite{LHW}. 

We do not have a complete understanding of the high value of 
$H_{c1}^*$ consistent with all experimental observations
at this time.  An obvious explanation is the existence of a 
surface barrier, as is commonly observed in type-II materials 
\cite{barrier}.  On the other hand, the expected hysteresis 
in such systems due to barrier profile asymmetry has apparently 
not been observed.

The real questions are as follows: For fields $H_{c1}<H< H_{c1}^*$, is 
the sample still in the pure Meissner state and can one 
expect to observe manifestations of the nonlinear Meissner 
effect? We believe not. Although the bulk magnetization has 
not occurred, the intermediate field state is thermodynamically 
unstable and vortices may be easily trapped at the sample 
corners or within the skin depth. Two recent high-resolution
penetration depth measurements \cite{Carrington,Bidinosti}
failed to measure the predicted temperature dependence in 
this field range, but did observe a large linear-$H$ field 
dependence most likely attributed to trapped vortices
\cite{Carrington}. These vortices may make the field 
penetration layer highly nonuniform, leading to 
a broadening 
of the $\breve{\rm Z}$uti\'c-Valls harmonics. 

Bhattacharya et al. argue \cite{goldmancomment,goldman}
that their failure to observe the
predicted small nonlinearities is strong evidence for the 
existence of a true gap in the {\it bulk} YBCO system, as 
created, e.g. by a bulk $d+is$ order parameter with 
$\Delta_s/\Delta_d$ of order a few percent. This claim is 
however inconsistent with several other experiments which 
indicate the existence of a residual normal fluid in $YBCO$, 
in one case down to 30mK \cite{tailleferuniversal}. Were a 
small out-of-phase order parameter component $\Delta_s$, of 
 order several degrees K, to exist, the gap in the density of
states would exponentially supress quasiparticle excitations 
at mK temperatures. On the other hand, a much smaller gap, 
of order tens or even hundreds of milliK is insufficient to 
eliminate the signal in the experiment of Ref. \cite{goldman}.  

The experiment of Bhattacharya et al.\cite{goldman} is a clever 
and careful attempt to measure a very small effect, and, as pointed 
out by $\breve{\rm Z}$uti\'c and Valls \cite{ZV}, in principle,  
is a sensitive probe of gap structure.  However, the interpretation
of the results is complicated by the influence of trapped flux, 
and it seems prudent that effects of this type  be ruled out or 
understood before claiming the observation of more exotic effects, 
such as the $d+is$ state invoked by Bhattacharya et al.
\cite{goldmancomment,goldman}.

\vskip .2cm
{\small M.-R. Li$^1$, P.J. Hirschfeld$^{2}$, and
P. W\"olfle$^1$
\vskip .01cm
\indent
$^1$Institut f\"ur Theorie der
Kondensierten Materie, Universit\"at Karlsruhe, 76128 Karlsruhe, 
Germany.
\vskip .01cm
\indent 
$^2$Department of Physics, University of 
Florida, 
Gainesville, FL 32611, USA.
\vskip .01cm

\vskip .2cm
\noindent Received  January 1999
}

\end{document}